\documentclass[aps,prb,twocolumn,showpacs,amsmath,amssymb,superscriptaddress,floatfix]{revtex4-1}
\usepackage{graphicx}
\usepackage{dcolumn}
\usepackage{bm}
\usepackage{amsfonts}
\usepackage{amsmath}
\usepackage{ulem}
\usepackage{color}
\begin{document}

\title{Full electrical control of Charge and Spin conductance \\through Interferometry of Edge States in Topological Insulators} 

\author{Fabrizio Dolcini}
\email{fabrizio.dolcini@polito.it}

\affiliation{Dipartimento di Fisica del Politecnico di Torino, I-10129 Torino, Italy}

\begin{abstract}
We investigate electron interferometry of edge states in Topological Insulators. We show that, when inter-boundary coupling is induced at two quantum point contacts of a four terminal setup, both  Fabry-P\'erot-like and Aharonov-Bohm-like loop processes arise. These underlying interference effects lead to a full electrically controllable system, where the magnitude of charge and spin linear conductances can be tuned by gate voltages, without applying magnetic fields. In particular we find that, under appropriate conditions, inter-boundary coupling can lead to negative values of the conductance. Furthermore, the setup also allows to selectively generate pure charge or pure spin currents, by choosing the voltage bias configuration.
\end{abstract}


\maketitle

%%%%%%%%%%%%%%%%%%%%%%%%%%%%%%%%%%%%%%%%%%%%%%%%%%%%%%%%%%%%%%%%%%%%%%%%%%%%%%%%%%%%
%%%%%%%%%%%%%%%%%%%%%%%%%%%%%%%%%%%%%%%%%%%%%%%%%%%%%%%%%%%%%%%%%%%%%%%%%%%%%%%%%%%%
%%%%%%%%%%%%%%%%%%%%%%%%%%%%%%%%%%%%%%%%%%%%%%%%%%%%%%%%%%%%%%%%%%%%%%%%%%%%%%%%%%%%
\section{Introduction}
Interferometry is a hallmark of wave physics. While optical interferometers have been known since past centuries, more recently  remarkable efforts have been made to realize  {\it electronic}  interferometers, which exploit  the wavelike nature of electrons, a signature of quantum world. Due to the decoherence effects that electrons experience  in ordinary macroscopic metals, such type of wavelike phenomenon can be observed only in mesoscopic systems. 
Fabry-P\'erot interference pattern caused by electron waves reflected at the contacts between a carbon nanostructure and two metallic electrodes have been observed in various experiments on nanotubes\cite{liang,recher,FP-other,bercioux} and graphene  \cite{FP-graphene}. Similarly, clear evidence of Aharonov-Bohm effect has been demonstrated in semiconductor heterostructures and graphene rings threaded by a magnetic flux\cite{AB-semi,AB-graphene}, and various electron interferometers have been realized in Quantum Hall Effect (QHE) systems.~\cite{heiblum,camino,west,rosenow}
 
Electron interferometry is not just a conceptually important effect, it has significant practical consequences, for it can be exploited to realize quantum transistors with high current-carrying capability\cite{FP-trans}. Indeed currents can be switched `on' and `off' by  varying the electron interference conditions from constructive to destructive through a gate voltage or a magnetic field.  \\

At present, most electronic interferometers are based on the electron  charge. However, electron is also characterized by its   spin, and spintronics -- the field  investigating transport and manipulation of information with such degree of freedom -- is experiencing an extremely  rapid  growth. Indeed spin is  much more robust to environment decoherence effects with respect to electron charge\cite{spintronics}, and spin   coherence lengths may reach and exceed  $100 \mu {\rm m}$.~\cite{long}  
For these reasons, interferometry involving spin has been proposed to realize transistors and  filters exploiting spin-orbit induced precession in semiconductors\cite{datta-das,nitta,marcus-gossard,frustaglia,xiao}, ferromagnetic 
materials\cite{schaepers,matsuyama,grundler,grifoni} and magnetic fields\cite{lundeberg}. In spite of these advances, interferometric control of spin currents  remains a difficult task requiring {\it ad hoc} optical techniques\cite{smirl,Zhao}, and has not reached the state of the art level.  Thus, as far as interferometry is concerned, spintronics is not as competitive as ordinary charge-based electronics yet.

A major boost to spintronics is expected to come from the recent discovery of Topological Insulators. A Topological Insulator (TI) is a bulk gapped material exhibiting conducting gapless channels at the boundaries.\cite{TI-reviews} In these edge states the generation of spin currents is greatly facilitated from the close connection between motion direction and spin orientation (helicity). In   two-dimensional realizations of a TI, for instance,  only spin-$\uparrow$ electrons propagate rightwards and only spin-$\downarrow$ electrons leftwards, along a given boundary. Remarkably, TI edge states behave as perfectly conducting one-dimensional ballistic channels, since impurity backscattering is prevented from time reversal symmetry. These peculiar properties, theoretically predicted\cite{kane-mele,zhang-PRL,zhang-science} and experimentally observed in HgTe/CdTe quantum wells\cite{molenkamp-science}  and in various other materials\cite{TI-other}, make TIs ideal candidates for  spintronics.\cite{TI-reviews,xiao,kim-qi,jiang,murakami} 
 
For these reasons the investigation of electron interference effects in these systems appears to be a particularly timely issue. Quite recently,  for instance, evidence of Aharonov-Bohm interference pattern in magnetoresistance of TIs has been experimentally observed\cite{peng} and theoretically discussed\cite{moore-PRL,kim-qi}. 
Notably, comparative analysis have pointed out that, with respect to the cases of edge states in  QHE  bars\cite{akhmerov} and SU(2) symmetric systems\cite{patrick}, interferometry of TI edge states exhibits intrinsically different behavior, and represents a challenging open problem with its own peculiarities.

So far, most studies concerning TI interferometry have involved magnetic flux or ferromagnets.\cite{peng,moore-PRL,kim-qi,akhmerov,patrick} However, TI edge states appear even in the absence of magnetic field, since the mechanism underlying their properties is the spin-orbit coupling and the related inversion of the electronic band order. Indeed this represents one of the crucial advantages of TIs with respect to QHE edge states in view of device miniaturization, due to the difficulty in realizing scalable circuits operating under high magnetic fields or space varying magnetizations with nanoscale resolution.   
It is thus desirable to explore TI interferometry also  in absence of magnetic fields or ferromagnets.

\begin{figure}[t]
\centering
%{setup-FP}
\includegraphics[width=0.95\linewidth]{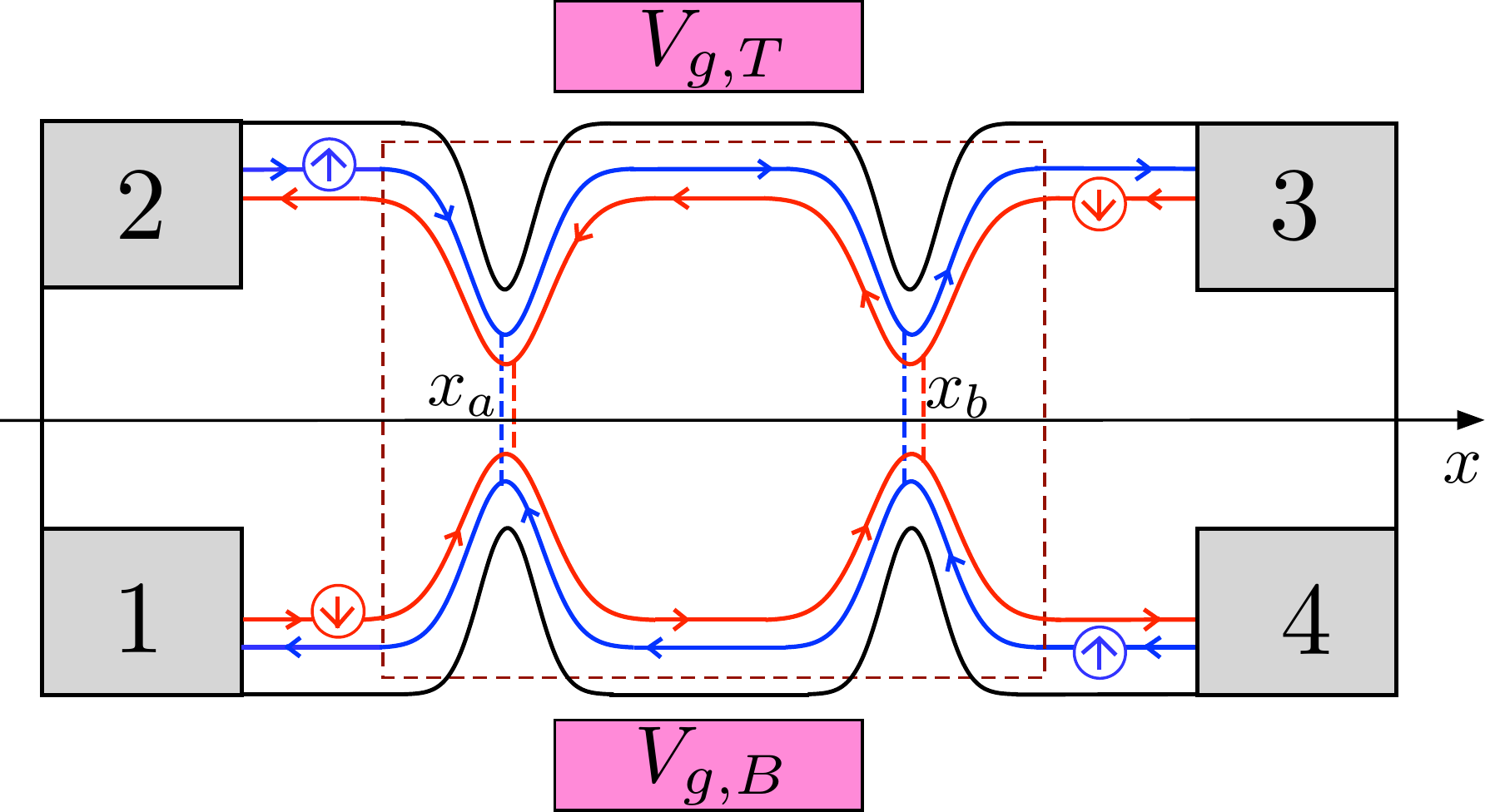}
\caption{\label{setup-Fig} (Color on line) Schematic description of the proposed four terminal setup, where edge states flow at the Top and Bottom boundaries of a TI quantum well. Blue (red) lines denote spin-$\uparrow$ ($\downarrow$) edge channels, and are slightly separated for illustrative reasons. The dashed box denotes the scattering region, where inter-boundary tunneling occurs at two QPCs, located around positions $x_a$ and $x_b$ and separated by a distance $L=x_b-x_a$.  Two gate voltages $V_{g,T}$ and $V_{g,B}$ shift the edge state momenta in the Top and Bottom regions between the QPCs, modifying the electron phase in the loop processes induced by tunneling. No magnetic flux is present.}
\end{figure}

In this article we address interferometry based on purely electrical effects, i.e. in the presence of time reversal symmetry. We shall focus on a two-dimensional realization of a TI, and consider the interferometer sketched in Fig.\ref{setup-Fig}, where Kramers pairs of TI edge states flow at the Top  and Bottom boundaries of a quantum well. The electrochemical potential of each injected edge state is controlled by  a metallic electrode, characterized by a voltage bias $V_i$ ($i=1,\ldots 4$). Inter-boundary scattering between edge states   can occur at  two Quantum Point Contacts (QPCs), giving rise to the possibility of loop trajectories, which determine the currents through electron wave interference.  
Here the  interference conditions are set by two (Top and Bottom) gate voltages $V_{g,T}$ and $V_{g,B}$, which modify  the electron phase in the loop processes by shifting the electron momenta in the regions between the two QPCs. The setup of Fig.\ref{setup-Fig} can thus lead to a full electrically controllable system, where charge and spin conductances of each electrode can be tuned by the gates. We shall determine and discuss the behavior of the conductances as a function of $V_{g,T}$ and $V_{g,B}$, for various configurations of the biases~$V_i$. In particular we shall show that, under appropriate conditions,  inter-boundary coupling can even lead to negative conductance values. Furthermore, the setup also allows to selectively generate and tune pure charge or pure spin currents.

The article is organized as follows. In Section~\ref{Sec2} we present the model for the setup, in Section~\ref{Sec3} we discuss the interference phenomena of the system and point out the relations with other electron interferometers. Then, in Section~\ref{Sec4} we show the results concerning charge and spin currents, mainly focussing on two specific configurations of the four terminal setup. Finally, we discuss the results in Section~\ref{Sec5} and draw our conclusions in Section~\ref{Sec6}.

%%%%%%%%%%%%%%%%%%%%%%%%%%%%%%%
%%%%%%%%%%%%%%%%%%%%%%%%%%%%%%%

%%%%%%%%%%%%%%%%%%%%%%%%%%%%%%%%%%%%%%%%%%%%%%%%%%%%%%%%%%%%%%%%%%%%%%%%%%%%%%%%%%%%%%%%%%%%%
%%%%%%%%%%%%%%%%%%%%%%%%%%%%%%%%%%%%%%%%%%%%%%%%%%%%%%%%%%%%%%%%%%%%%%%%%%%%%%%%%%%%%%%%%%%%%
%%%%%%%%%%%%%%%%%%%%%%%%%%%%%%%%%%%%%%%%%%%%%%%%%%%%%%%%%%%%%%%%%%%%%%%%%%%%%%%%%%%%%%%%%%%%%
%%%%%%%%%%%%%%%%%%%%%%%%%%%%%%%%%%%%%%%%%%%%%%%%%%%%%%%%%%%%%%%%%%%%%%%%%%%%%%%%%%%%%%%%%%%%%
%%%%%%%%%%%%%%%%%%%%%%%%%%%%%%%%%%%%%%%%%%%%%%%%%%%%%%%%%%%%%%%%%%%%%%%%%%%%%%%%%%%%%%%%%%%%%
%%%%%%%%%%%%%%%%%%%%%%%%%%%%%%%%%%%%%%%%%%%%%%%%%%%%%%%%%%%%%%%%%%%%%%%%%%%%%%%%%%%%%%%%%%%%%
%%%%%%%%%%%%%%%%%%%%%%%%%%%%%%%%%%%%%%%%%%%%%%%%%%%%%%%%%%%%%%%%%%%%%%%%%%%%%%%%%%%%%%%%%%%%%
\section{The model}
\label{Sec2}
The edge states at the Top and Bottom boundaries of the device  are characterized, at low energies, by a linear spectrum  described by the following Hamiltonian\cite{chamon_2009}
\begin{eqnarray}
{\mathcal{H}}_0  &=&   -i    \hbar v_{\rm F} \sum_{\sigma=\uparrow, \downarrow}\! \int   \! dx  \left[:   \Psi^{\dagger}_{R \sigma}(x)\, \partial_x \Psi^{}_{R \sigma}(x):  \right. \nonumber \\
& & \hspace{2.5cm} \left. -  :\Psi^{\dagger}_{L\bar{\sigma}}(x) \, \partial_x \Psi^{}_{L \bar{\sigma}}(x) \, :\right] \,           
\label{H0W}
\end{eqnarray}
where $x$ denotes the longitudinal coordinate,  $\Psi_{R \sigma}$ and $\Psi_{L \sigma}$ the Right and Left mover electron field operators, and $\sigma=\uparrow,\downarrow$ the spin component. Here we adopt the notation  that  spin-$\uparrow$ Right (Left) movers and spin-$\downarrow$ Left (Right) movers flow along the Top (Bottom) boundary, as depicted in Fig.\ref{setup-Fig}.  Without loss of generality we shall assume that the equilibrium Fermi level $E_F$ of the device in the absence of any bias is located at the Dirac point of the spectrum. The symbol  $: \, \, \, :$ in Eq.(\ref{H0W}) denotes the normal ordering with respect to the equilibrium state where all levels below $E_F$ are occupied. Variations from this energy level can be induced by the gate voltages, as discussed below.

The constrictions of the QPCs induce inter-boundary scattering. It can be shown with quite general arguments\cite{zhang-PRL,teo} that time-reversal symmetry only allows  two types of tunneling terms, namely a spin-preserving tunneling
\begin{eqnarray}
{\mathcal{H}}^{p}_{tun} &=&    \sum_{\sigma=\uparrow, \downarrow}\! \int   \! dx      \left( \, \Gamma_{p}(x)  \,   \Psi^{\dagger}_{R \sigma}(x)\, \Psi^{}_{L \sigma}(x)  \,+ \, \right. \nonumber \\
& & \hspace{1.5cm}  \left. +\Gamma_{p}^{*}(x) \,  \Psi^{\dagger}_{L \bar{\sigma}}(x)\, \Psi^{}_{R \bar{\sigma}}(x) \right) \,  
\label{Htun-sp} 
\end{eqnarray}
and a spin-flipping tunneling
\begin{eqnarray}
{\mathcal{H}}^{f}_{tun} &= &  \! \! \sum_{\alpha=R\!/\!L=\pm}  \alpha \! \int   \! dx    \,  \left( \, \Gamma_{f}(x) \,    \Psi^{\dagger}_{\alpha \uparrow}(x)\, \Psi^{}_{\alpha \downarrow}(x)  \,+ \right. \nonumber \\
& & \hspace{2.2cm} \left. \, +\Gamma_{f}^*(x) \, \,  \Psi^{\dagger}_{\alpha \downarrow}(x)\, \Psi^{}_{\alpha \uparrow}(x)  \right) \,  \;  
\label{Htun-sf} 
\end{eqnarray}
In fact, the pinching of the two edges states at the QPCs causes a  local  modification of the spin-orbit coupling with respect to the bulk case, so that both terms are expected to contribute.\cite{teo,trauz-recher}
In Eqs.(\ref{Htun-sp}) and (\ref{Htun-sf})  $\Gamma_{p,f}(x)$ denote    space-dependent tunneling  amplitude
profiles. The two QPCs will thus be described through a profile  peaked around two centers $x_a$ and $x_b$ (see Fig.\ref{setup-Fig}), and rapidly decaying beyond a longitudinal lengthscale $\xi$. We shall assume that the constriction are short with respect to the Fermi wavelength~$\lambda_F$, and that   the distance $L=x_b-x_a$ between the two QPCs is large compared to $\xi$, i.e. $\xi \lesssim \lambda_F < L$. Under these conditions, it is sufficient to assume that tunneling is point-like. 
For the sake of simplicity we shall also consider that the two QPCs are characterized by equal tunneling amplitudes, so that the profiles can be taken as 
\begin{equation}
\Gamma_{p(f)}(x)= 2 \hbar v_{F} \, \gamma_{p(f)}\sum_{l=a,b}  \, \delta(x-x_l) 
\end{equation} 
where $\gamma_{p}$ and $\gamma_{f}$ are real dimensionless spin-preserving and spin-flipping tunneling amplitudes, respectively.\\

Finally, the coupling to the two gate voltages $V_{g,T}$ and~$V_{g,B}$ can be described by the term
\begin{eqnarray}
{\mathcal{H}}_{g} &=& \! \!   \int_{x_a}^{x_b}   \!  \! dx    \left[ \, eV_{g,T} \left( \, \rho^{}_{R \uparrow}(x) \, + \rho^{}_{L \downarrow}(x)  \right)  + \right. \label{Hgate}\\
&   & \hspace{0.7cm} + \, \left. eV_{g,B} \left( \,\rho^{}_{R \downarrow}(x)\, +  \rho^{}_{L \uparrow}(x)  \right) \right]
\nonumber 
\end{eqnarray}
where  
\begin{equation}
\rho_{\alpha \sigma} \doteq \, : \Psi^\dagger_{\alpha \sigma}(x) \Psi^{}_{\alpha \sigma}(x) : \label{rho-def}
\end{equation}
denotes the electron density, with $\alpha=R/L$ and $\sigma=\uparrow,\downarrow$. As anticipated above, $V_{g,T}$ and $V_{g,B}$ shift the electronic spectrum (\ref{H0W}), and their difference  breaks the degeneracy between Top and Bottom boundaries.\\

The equations of motion for the four fields $\Psi_{\alpha \sigma}$  are easily obtained from the total Hamiltonian of the device  
\begin{equation}
{\mathcal{H}} ={\mathcal{H}}_{0}     \, + \,       
  {\mathcal{H}}^{p}_{tun} \, + \,   {\mathcal{H}}^{f}_{tun} \, +    
  {\mathcal{H}}_{g}\;   \label{HAM-FER}
\end{equation}
where the four terms are given in Eqs.(\ref{H0W}), (\ref{Htun-sp}), (\ref{Htun-sf}) and~(\ref{Hgate}), respectively. For $\Psi^{}_{R \uparrow}$, for instance, one obtains
\begin{eqnarray}
\partial_t \Psi^{}_{R \uparrow}   & =& \,v_{\rm F} \left( -    \, \partial_x \Psi_{R \uparrow} - i k_{g,R \uparrow} \theta(x_a < x< x_b) \, \Psi^{}_{R \uparrow}(x)-\right. \nonumber \\
& & \left. \hspace{-0.7cm} -2 i \sum_{l=a,b} 
\left( \gamma_{p}  \, \Psi_{L \uparrow}(x) \, + \gamma_{f}  \, \Psi_{R \downarrow}(x) \right) \delta(x-x_l)   \, \right) 
\end{eqnarray}
where $\theta$ is the Heaviside function. Similar equations are obtained for the other fields. The four equations, coupled by the inter-boundary tunneling terms,   can be solved by superposing  solutions corresponding to fixed energy values. At a given energy~$E$, measured with respect to the equilibrium Fermi level $E_F$, the solution can be obtained by the Ansatz
\begin{eqnarray}
\Psi_{E R \sigma}(x) &=& \frac{e^{-\frac{i}{\hbar} E t}}{\sqrt{h v_{\rm F}}}  \left\{ 
\begin{array}{lcl} 
\hat{a}_{E R \sigma} \, e^{i k_E x} & & \mbox{$x<x_a$} \\
\hat{c}_{E R \sigma} \, e^{i (k_E-k_{g,R \sigma}) x} & & \mbox{$x_a<x<x_b$} \\
\hat{b}_{E R \sigma} \, e^{i k_E x} & & \mbox{$x>x_b$} \\
\end{array}
\right. \nonumber  \\ & & \label{Ansatz} \\
\Psi_{E L \sigma}(x) &=& \frac{ e^{-\frac{i}{\hbar} E t} }{\sqrt{h v_{\rm F}}}\left\{ 
\begin{array}{lcl} 
\hat{b}_{E L \sigma} \, e^{-i k_E x} & & \mbox{$x<x_a$} \\
\hat{c}_{E L \sigma} \, e^{-i (k_E-k_{g,L \sigma}) x} & & \mbox{$x_a<x<x_b$} \\
\hat{a}_{E L \sigma} \, e^{-i k_E x} & & \mbox{$x>x_b$} \\
\end{array}
\right. \nonumber
\end{eqnarray}
where $k_E=E/\hbar v_{F}$, $k_{g, R \uparrow}=k_{g, L \downarrow} =e V_{g, T}/\hbar v_F$, $k_{g, R \downarrow} =k_{g, L \uparrow}=e V_{g, B}/\hbar v_F$, and $\hat{c}_{E \alpha \sigma}$ denotes the electron operator inside the scattering region. The incoming and outgoing electrons, respectively described   by the operators $\hat{a}$ and $\hat{b}$ in (\ref{Ansatz}),  are connected via the entries $S_{ij}$ of the Scattering Matrix $\mathsf{S}$~\cite{butti-multi}
\begin{eqnarray}
\left(
\begin{array}{c}
{\hat{b}}^{}_{E L \uparrow} \,  \\ \\    
{\hat{b}}^{}_{E L \downarrow} \, \\ \\
{\hat{b}}^{}_{E R \uparrow} \,  \\ \\ 
{\hat{b}}^{}_{E R \downarrow} \, 
\end{array}
\right)  
= \left( 
\begin{array}{cccc} 
S_{11} & S_{12} & S_{13} & S_{14} \\ \\
S_{21} & S_{22} & S_{23} & S_{24} \\ \\
S_{31} & S_{32} & S_{33} & S_{34}\\ \\
S_{41} & S_{42} & S_{43} & S_{44} \end{array}
\right)  \cdot \left(
\begin{array}{c}
{\hat{a}}^{}_{E R \downarrow}   \\ \\   {\hat{a}}^{}_{E  R \uparrow} \, \\ \\
{\hat{a}}^{}_{E L \downarrow}   \\ \\   {\hat{a}}^{}_{E  L \uparrow} \,   
\end{array}
\right)  \label{S-FP}
\end{eqnarray}
In the columns of Eq.(\ref{S-FP})  $\hat{a}$ and $\hat{b}$ operators appear according to the clockwise order of the four leads in the setup Fig.\ref{setup-Fig}.
A lengthy but straightforward calculation allows to determine the Scattering Matrix   entries $S_{ij}$ as a function of the tunneling amplitudes $\gamma_{p}, \gamma_{f}$ and the gate voltages $V_{g,T}$ and $V_{g,B}$. 
Helicity and time-reversal symmetry lead to vanishing diagonal entries,
\begin{equation}
S_{ii} = 0 \quad,
\end{equation}
while the other entries can be expressed in a compact form by utilizing the transmission coefficients of each single QPC in terms of the tunneling amplitudes $\gamma_p$ and $\gamma_f$,
\begin{eqnarray}
T^{a}_{21} = T^{a}_{34} &= & \frac{4 \gamma_{p}^2}{(1+\gamma_p^2+\gamma_f^2)^2} \nonumber \\
T^{a}_{31} = T^{a}_{42}    &= & \frac{4 \gamma_{f}^2}{(1+\gamma_p^2+\gamma_f^2)^2}  \label{Tij1-def} \\
T^{a}_{41} =T^{a}_{32}  &= & \frac{(1-\gamma_p^2-\gamma_f^2)^2}{(1+\gamma_p^2+\gamma_f^2)^2} \nonumber \quad .  
\end{eqnarray}
Here $T^{a}_{ij}$ ($T^{b}_{ij}$) denotes the transmission coefficient from lead $j$ to lead $i$ of the Left (Right) QPC alone. Time reversal symmetry ensures
\begin{equation}
T^{l}_{ij}=T^{l}_{ji} \hspace{1cm} l=a,b \quad.
\end{equation} 
Furthermore, we have assumed that the QPCs have identical parameters
\begin{equation}
T^{a}_{ij}=T^{b}_{ij}  \quad .
\end{equation} 
Then, the ${\mathsf{S}}$-matrix entries read 
%\begin{widetext}
\begin{eqnarray}
%|S_{12}(E)|^2 &=&\frac{8 \gamma_{p}^2 \, (1+\gamma_{p}^2+\gamma_{f}^2)^2 \, (1+\cos[2(k_E-\frac{k_{g, T}+k_{g, B}}{2}) L])}{(1+\gamma_{p}^2+\gamma_{f}^2)^4+16 \gamma_{p}^4+8 \gamma_{p}^2 (1+\gamma_{p}^2+\gamma_{f}^2)^2 \cos{[2(k_E-\frac{k_{g,B} +k_{g,T}}{2})L]}}
%\\ & & \nonumber \\
%|S_{13}(E)|^2 &=&\frac{8 \, \gamma_{f}^2  \, (1-\gamma_{p}^2-\gamma_{f}^2)^2 \,(1+\cos{[ (k_{g, T}-k_{g, B})L ]})}{(1+\gamma_{p}^2+\gamma_{f}^2)^4+16 \gamma_{p}^4+8 \gamma_{p}^2 (1+\gamma_{p}^2+\gamma_{f}^2)^2 \cos{[2(k_E-\frac{k_{g,B} +k_{g,T}}{2})L]}} 
%\\ & & \nonumber \\
%|S_{14}(E)|^2 &=&\frac{(1-\gamma_{p}^2-\gamma_{f}^2)^4+16 \gamma_{f}^4-8 \gamma_{f}^2 (1-\gamma_{p}^2-\gamma_{f}^2)^2 \cos[{(k_{g,T} -k_{g,B})L}]}{(1+\gamma_{p}^2+\gamma_{f}^2)^4+16 \gamma_{p}^4+8 \gamma_{p}^2 (1+\gamma_{p}^2+\gamma_{f}^2)^2 \cos{[2(k_E-\frac{k_{g,B} +k_{g,T}}{2})L]}} \\
%%%%%
S_{31} &= & -i  \,  \frac{\sqrt{T^{a}_{31} T^{b}_{32}}   \, e^{-i k_{g, T}  L} +\sqrt{T^{a}_{41} T^{b}_{31} } \, e^{-i k_{g, B}  L}  }
{1+\sqrt{T^{a}_{43} T^{b}_{21}} \, \, e^{2i (k_E-\frac{k_{g,T} +k_{g,B}}{2})L}} \nonumber \\ & & \nonumber \\
%%%%
S_{32}&=&  \frac{\sqrt{T^{a}_{32} T^{b}_{32}} e^{-ik_{g,T} L}-\sqrt{T^{a}_{42} T^{b}_{31}}  e^{-ik_{g,B} L}}{1+\sqrt{T^{a}_{43} T^{b}_{21}} \, \, e^{2i(k_E-\frac{k_{g,T} +k_{g,B}}{2})L}} \label{Smatr-entries} \\ & & \nonumber \\
%%%%
S_{34} &= &- i \, e^{-2 i k_E x_b} \, \frac{ \sqrt{T^{b}_{34}} \, +\sqrt{T^{a}_{34}}\, \, e^{2i(k_E-\frac{k_{g,B} +k_{g,T}}{2})L}}{1+\sqrt{T^{a}_{43} T^{b}_{21}} \, \, e^{2i(k_E-\frac{k_{g,T} +k_{g,B}}{2})L}}  \nonumber
%%%%
\end{eqnarray} 
where
\begin{equation}
k_{g,T/B} = \frac{e V_{g,T/B}}{\hbar v_F} \quad. \label{kgTB}
\end{equation}
Similar expressions are found for the other entries, together with the following relations
\begin{eqnarray}
|S_{13}(E)| =|S_{31}(E)|=|S_{24}(E)|=|S_{42}(E)| \nonumber \\ \label{S-prop} \\
|S_{14}(E)| =|S_{41}(E)|=|S_{23}(E)|=|S_{32}(E)| \nonumber
\end{eqnarray}
The $\mathsf{S}$ matrix entries  allow to define the system transmission coefficient   from lead $j$ to lead $i$ (computed at the Fermi energy, i.e. at $E=0$)  as
\begin{equation}
T_{ij} \doteq |S_{ij}(0)|^2 \label{Tij-def}
\end{equation}
It is well known\cite{butti-multi} that $T_{ij}$ appear in the expressions for the current, as will be explicitly discussed below.

\section{Interference phenomena \\in the system}
\label{Sec3}
Before presenting the results for the currents (see Section \ref{Sec4}), in this section we describe the interference effects characterizing the setup. This gives us the opportunity to point out  similarities and   differences with respect to other electron interferometers.
\subsection{Fabry-P\'erot interference}
\label{Sec-FP}
To simplify the discussion let us first assume that the spin-flipping tunneling amplitude is vanishing ($\gamma_{f}=0$). In this case quantum interference is due to spin-preserving processes ($\gamma_{p} \neq 0$). An illustrative example is described in Fig.\ref{interf}(a): a spin-$\uparrow$ electron is injected from  terminal 2 and, when reaching the right QPC,  partly  tunnels  to the Bottom boundary as a left-mover; it then tunnels back to the Top boundary at the left QPC and eventually reaches terminal 3.   The phase accumulated in the loop causes the interference with the electron wave that straightforwardly travels from terminal 2 to 3. Similar processes occur for spin-$\downarrow$ electrons  injected from  terminal 1 to 4. 
Notice that these processes involve a change in the motion direction, i.e. an energy dependent  momentum transfer $(e^{i k_E x} \rightarrow e^{-i k_E x})$, with $k_E=E/\hbar v_F$. \\

This effect is reminiscent of the Fabry-P\'erot (FP) interference pattern occurring in carbon nanotubes\cite{liang,recher,balents} or in single channel quantum wires\cite{pugnetti_2009}, and involving $2 k_{F}$ momentum transfer.  
Indeed the inter-boundary terms (\ref{Htun-sp})  causing the loop of Fig.\ref{interf}(a)  correspond to backscattering terms originating from  the contact resistance at the  nanotube(wire)/electrode  interfaces, while  the distance $L$ between the two QPCs plays the role of the length of the nanotube (wire).

The electron phase  accumulated in the loop depends on the gate voltages, which modify the momenta in the Top and Bottom region between the QPCs. In view of the analogy discussed above, we shall denote such phase
\begin{equation}
\phi_{FP}=\frac{e(V_{g,T}+V_{g,B}) L}{\hbar v_F} \label{FP-phase}
\end{equation}
as the Fabry-P\'erot (FP) phase of the device. 
Notice that~$\phi_{FP}$ depends on the {\it sum} $V_{g,B} +V_{g,T}$ of the gate voltages   and on the QPC distance $L$.

%%%%%%%%%%%%%%%%%%%%%%%%%%%%%%%%%%%%%%%%%%%%%%%%%%
\subsection{Aharonov-Bohm interference}
\label{Sec3B}
\begin{figure}[h]
\centering
%{interference-processes}
\includegraphics[width=0.8\linewidth]{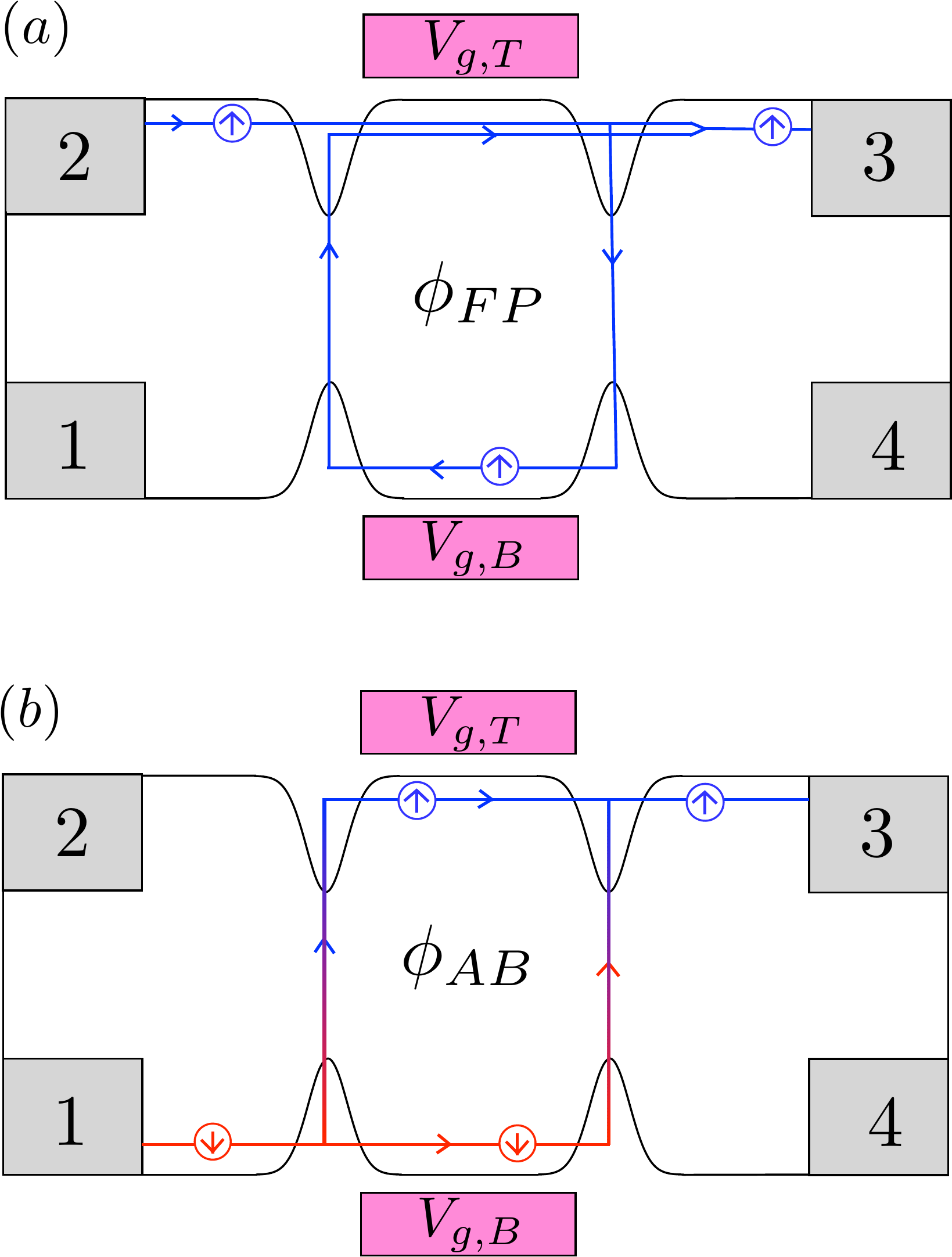}
\caption{\label{interf} The two types of basic interference processes allowed by time reversal symmetry  of the system. (a) An example of loop process  induced  by the spin-preserving tunneling terms Eq.~(\ref{Htun-sp}), reminiscent  of a Fabry-P\'erot interferometer. The electron flows rightwards in the Top Boundary and leftwards in the Bottom boundary, so that the loop interference phase is $\phi_{FP}=e(V_{g,T}+V_{g,B}) L/\hbar v_F$.  \\(b) An example of loop processes induced   by  the spin-flipping tunneling terms Eq.~(\ref{Htun-sf}), reminiscent of electrostatic Aharonov-Bohm interferometer. The electron maintains the motion direction both in the Top Boundary and in the Bottom boundary, so that the loop interference phase is  $\phi_{AB}=e(V_{g,T}-V_{g,B}) L/\hbar v_F$. A similar loop connects terminal 2 to~3.\\ 
In general multi-loop processes, where the two processes interplay, are also possible.}
\end{figure}

Let us now consider the effect of the spin-flipping terms ($\gamma_{f} \neq 0$), and assume that the spin-preserving processes are absent ($\gamma_{p}=0$). Interference processes for this situation are schematically depicted in Fig.\ref{interf}(b): A right-moving spin-$\downarrow$ electron  injected from  terminal 1 exhibits two possible paths to reach terminal 3, corresponding to   tunneling to the Top boundary occurring at the   QPC on the left and on the right.  
The phase difference  
\begin{equation}
\phi_{AB}=\frac{e(V_{g,T}-V_{g,B}) L}{\hbar v_F} \label{AB-phase}
\end{equation}
between these two possible paths connecting terminals 1 and~3 causes an electron interference 
phenomenon similar to  the electrostatic Aharonov-Bohm (AB) effect occurring in semiconductor and metallic rings rings.\cite{AB-electro-theo,AB-electro-exp}  Similar AB-like loop trajectories connect terminals~2 and~3.
Notice that, differently from the spin-preserving FP loop processes, the AB interference due to spin-flipping processes is independent of the energy~$E$ and involves the {\it difference} $V_{g,T}-V_{g,B}$ between the two gate voltages. This is due to the fact that, while in FP processes the electron travels along the two arms of the loop in opposite directions [see Fig.\ref{interf}(a)],  for spin-flipping processes the electron always preserves its motion direction in order for the spin to flip [see Fig.\ref{interf}(b)]. This constraint originates from time reversal symmetry, which leads spin-preserving inter-boundary tunneling to occur backwards [$R \leftrightarrow L$, see Eq.(\ref{Htun-sp})], and spin-flipping inter-boundary tunneling to occur forwards [$\alpha \rightarrow \alpha$, see Eq.(\ref{Htun-sf})]. This connection between spin orientation and motion direction   is   a hallmark of topological insulator edge states dynamics.

%%%%%%%%%%%%%%%%%%%%%
%%%%%%%%%%%%%%%%%%%%% 
\subsection{General case}
\label{Sec3C}
The two processes of Fig.\ref{interf} described above refer to the extreme cases where either $\gamma_p$ or $\gamma_f$ vanishes. 
In the general case ($\gamma_p \neq 0$ and $\gamma_f \neq 0$) higher order processes give rise to multiple loops where the Fabry-P\'erot  and the Aharonov-Bohm interference effects interplay with each other. These processes directly enter the Scattering Matrix entries (\ref{Smatr-entries}) and thus the transmission coefficients (\ref{Tij-def}).
In particular, to lowest order in the  tunneling amplitudes $\gamma_p$ and $\gamma_f$, the `horizontal' transmission  $T_{32}$ is determined by both Fabry-P\'erot  and Aharonov-Bohm interference processes, whereas the `crossing' transmission $T_{31}$ is only affected by Aharonov-Bohm loops. Indeed, while terminal 2 can be connected to terminal 3  through either one FP loop or one AB loop (see Fig.\ref{interf}), terminals 1 and 3 can only by connected  through one AB loop, as can be formally checked by   perturbative expansion of  Eqs.(\ref{Smatr-entries}) and~(\ref{Tij-def}).

\subsection{Differences from other electron interferometers.}
In the previous subsections we have highlighted  the analogies between the basic interference phenomena of the setup Fig.\ref{setup-Fig} and other electron interferometers. Here we wish to  point  out the differences. \\

The first difference  emerges  at the level of the coupling to the gate. To illustrate this point, one can realize that the gate coupling term (\ref{Hgate}) can be rewritten in the following way
\begin{eqnarray}
{\mathcal{H}}_{g} &=& \! \!   \int_{x_a}^{x_b}   \! \!  \! dx    \left[ \, (V_{g,T}+V_{g,B})   \, \rho^{}_{c}(x) \, +     \right. \label{Hgate-bis}\\
&   &     \,   + \left.  v_F^{-1} (V_{g,T}-V_{g,B})    \ I^{}_{s}(x)  \right]
\nonumber 
\end{eqnarray}
where
\begin{eqnarray}
\rho_{c}(x) &=& e \, \langle  \rho_{R \uparrow}(x)+\rho_{L \uparrow}(x)+  \rho_{R \downarrow}(x)+\rho_{L \downarrow}(x)  \rangle \label{rhoc}
\end{eqnarray}
is the charge density operator, and 
\begin{eqnarray}
I_s(x) &=& e v_{\rm F}  \, \langle     \rho_{R \uparrow}(x)   - \rho_{L \uparrow}(x)  -  \rho_{R \downarrow}(x) +\rho_{L \downarrow}(x) \rangle \hspace{0.5cm} \label{Is-def}  
\end{eqnarray}
is the spin current operator\cite{nota-1}, with the  electron densities $\rho_{\alpha \sigma}$ given by~Eq.(\ref{rho-def}).
Equation (\ref{Hgate-bis}) shows that the sum of the gate voltages, i.e. the FP phase~(\ref{FP-phase}), couples to the  charge density, while their difference, i.e. the AB  phase~(\ref{AB-phase}), couples to the spin current. The expression (\ref{Hgate-bis}) points out a difference with respect to Fabry-P\'erot electron interferometers realized with carbon nanotubes or quantum wires. 
In these systems spin-$\uparrow$ and spin-$\downarrow$  electrons flow along the same physical channel; the gate voltage applies in the same way to all $\rho_{\alpha \sigma}$ ($\alpha=R/L$ and $\sigma=\uparrow,\downarrow$), so that it effectively couples to the charge density $\rho_c(x)$ only. In contrast, in TIs  the space separation between boundaries allows for the presence of two different gate voltages [see Eq.(\ref{Hgate})], and the helical nature of the edge states  leads to the coupling to the spin current $I_s(x)$ as well. 
As we shall see, this additional handle offers interesting possibilities to control the spin current. \\

The fact that only $\rho^{}_{c}$ and $I^{}_{s}$ appear in Eq.(\ref{Hgate-bis})   is due to the time-reversal symmetry of the system. Indeed  electron charge and spin current operators   are the only two combinations of the four densities $\rho_{\alpha \sigma}$ that are even under time reversal, whereas the electron current 
\begin{eqnarray}
I_{c}(x) &=& e v_{\rm F} \, \langle  \rho_{R \uparrow}(x)-\rho_{L \uparrow}(x)+  \rho_{R \downarrow}(x)-\rho_{L \downarrow}(x)  \rangle \hspace{0.5cm} \label{Ic-def} 
\end{eqnarray}
and spin density 
\begin{eqnarray}
\rho_s(x) &=& e \, \langle     \rho_{R \uparrow}(x)   + \rho_{L \uparrow}(x)  -  \rho_{R \downarrow}(x) -\rho_{L \downarrow}(x) \rangle \hspace{0.5cm} \label{rhos-def}  
\end{eqnarray}
are odd. 
This enables us to  highlight  the differences with respect to the more traditional magnetic Aharonov-Bohm effect investigated in semiconductors\cite{AB-semi} and graphene\cite{AB-graphene} rings,  QHE systems\cite{west,rosenow}, and also  TI edge states\cite{patrick}. In these systems a coupling occurs between the vector potential and the charge current $I_c$, and the interference is driven by the magnetic flux, breaking time reversal symmetry. In contrast, the Aharonov-Bohm phase (\ref{AB-phase}) related to the process in Fig.\ref{interf}(b) originates from a purely {\it electrostatic} effect, preserving time-reversal symmetry: the electron momenta in the two `arms' of the ring are different whenever $V_{g,T}-V_{g,B} \neq 0$.  

On the other hand, the interferometric process in Fig.\ref{interf}(b) also differs from the electrostatic Aharonov-Bohm effect\cite{AB-electro-theo,AB-electro-exp}. While in such effect the electron spin plays no role,   here the helical properties of TI edge states relate the AB phase $\phi_{AB}$ to spin flipping processes, thus modifying the spin current, as we shall describe below. 
A distinction can also be pointed out with respect to the Aharonov-Casher effect, observed in various mesoscopic semiconductor ring nanostructures\cite{nitta,AC}. In that case a spin precession along the ring arms is caused by a uniform Rashba spin-orbit coupling, which can be controlled via a gate by varying the asymmetry of the quantum well structure. This yields a spin-dependent phase difference between the two ring arms\cite{ihn-book}. In contrast, here spin flip processes occur only locally   at the QPCs, and $\phi_{AB}$ is actually independent of the spin.  \\

Finally, another important difference with respect to   electron interferometers realized with nanotubes or semiconductor rings is concerned with the coupling to the biasing electrodes. Such interferometers consist of two-terminal setups, where spin-$\uparrow$ and spin-$\downarrow$ electrons are injected from the same   electrode and flow along the same physical channel. In contrast, TI edge states are geometrically separated and the four terminals  in Fig.\ref{setup-Fig} allow  for an independent control  of the four injected electron species. For these reasons the topology of the interferometer under investigation here is intrinsically much richer than the above interferometers. As we shall show, this property enables a tunability of the type (charge or spin), the magnitude and, in some cases, also of the sign of the currents.   
%%%%%%%%%%%%%%%%%%%%%%%%%%%%%%%%%%%%%%%%%%%%%%%%%%%%%%%%%%%%%%
%%%%%%%%%%%%%%%%%%%%%%%%%%%%%%%%%%%%%%%%%%%%%%%%%%%%%%%%%%%%%%
%%%%%%%%%%%%%%%%%%%%%%%%%%%%%%%%%%%%%%%%%%%%%%%%%%%%%%%%%%%%%%
%%%%%%%%%%%%%%%%%%%%%%%%%%%%%%%%%%%%%%%%%%%%%%%%%%%%%%%%%%%%%%
%%%%%%%%%%%%%%%%%%%%%%%%%%%%%%%%%%%%%%%%%%%%%%%%%%%%%%%%%%%%%%
\section{Results for Currents}
\label{Sec4} 
We shall now present the results concerning the currents   in terms of the applied biases and  gate voltages. The definitions of charge and spin current  are given by Eqs.(\ref{Ic-def}) and (\ref{Is-def}), respectively.  
Using Eq.~(\ref{Ansatz}) and   exploiting the Scattering matrix (\ref{S-FP}), the   currents flowing in each terminal are easily evaluated in terms of the Fermi distributions $f_j(E)$ ($j=1,\ldots 4$) of the  states injected from the four leads\cite{butti-multi}. Due to the presence of four distribution functions (each including bias voltage and  temperature) and to   various parameters of the Fabry-P\'erot interferometer,  the behavior of the current exhibits an extremely rich scenario. Since we are interested in the quantum regime, we shall limit our analysis to the  zero temperature case. Furthermore, we shall focus on configurations  where  non vanishing voltage biases are applied to the electrodes on the left-hand side of  the sample (terminals 1 and 2), and the currents are measured in the terminals on the right-hand side   (3 and 4), which are assumed to be grounded, i.e. $f_{3}=f_{4}=f_{eq}$, with $f_{eq}$ denoting the Fermi distribution of any lead at equilibrium. Under these circumstances, the expression for the charge current in the $i$-th lead ($i=3,4$) reduces to 
\begin{eqnarray}
I_c(i)=\frac{e}{h} \int   dE    \sum_{j=1,2}  |S_{i j}(E)|^2   (f_{j}(E)-f_{eq}(E))  \hspace{0.5cm} \label{Ic-gen}
\end{eqnarray}
and    the spin currents fulfill the following relations
\begin{equation}
\left\{ 
\begin{array}{lcl} 
I_s(3)&=& I_c(3) \\ & & \\ I_s(4)&=& -I_c(4) \quad.
\end{array}
\right. \label{Is-gen}
\end{equation}
Equation (\ref{Is-gen}) points out that the charge currents measured in terminals 3 and 4 are always accompanied by a spin current, as expected from the peculiar properties of TI edge states. The difference in the relative sign between charge and spin currents can be easily understood even in the absence of the QPCs.
Indeed a positive voltage bias applied to terminal 2   injects spin-$\uparrow$ electrons into the scattering region, whereas a positive voltage bias applied to terminal 1 injects  spin-$\downarrow$ electrons, leading to parallel charge currents   and   counter-flowing spin currents [see Eqs.(\ref{Is-def}) and (\ref{Ic-def})]. \\ 
In particular, we are interested in two configurations of applied voltage biases $V_1$ and $V_2$, namely
\begin{eqnarray}
\begin{array}{ll}
\mbox{(C)} & \mbox{\underline{`Charge'-bias }}  \\
& \mbox{\underline{Configuration}} 
\end{array}
&=&  
\left\{ 
\begin{array}{lcl} 
V_{2}&=& V_{1}=V \\ & & \\
V_{3}&=& V_{4}=0 
\end{array} \right. \hspace{0.5cm} \label{conf-c} 
\\ & & \nonumber \\ & & \nonumber \\
\begin{array}{ll}
\mbox{(S)} & \mbox{\underline{`Spin'-bias }}  \\
& \mbox{\underline{Configuration}} 
\end{array}
&=&  
\left\{ 
\begin{array}{lcl}   
V_{2}&=&-V_{1}=V \\ & & \\ V_{3}&=&V_{4}=0 \end{array}
\right. \hspace{0.5cm}\label{conf-s}
\end{eqnarray}

The labels `Charge' and `Spin' associated with these bias configurations originate from the overall degree of freedom injected into the scattering region, depicted in Fig.\ref{setup-Fig} as a dashed box. 
Indeed in configuration~(C) the amount of spin-$\uparrow$ and spin-$\downarrow$ electrons injected  from terminals 1 and 2 is the same, so that effectively only the charge degree of freedom is injected, and no net spin. In contrast, in configuration~(S) the lead 1 is negatively biased, determining a depletion of   spin-$\downarrow$ electrons with respect to the equilibrium situation. In this case only a spin degree of freedom is supplied to the scattering region, with no net amount of injected charge. Similar configurations have been discussed in 
Ref.[\onlinecite{trauz-recher}].
In intermediate situations, i.e. when $|V_1| \neq |V_2|$, both charge and spin degree of freedom are involved. The injected degree of freedom (charge and/or spin) experiences scattering events due to the inter-boundary tunneling terms, which  determine the current actually measured in terminals 3 and 4. For each configuration  one can  define charge and spin conductances in the $i$-th lead as the linear response of $I_c$ and $I_s$  to the bias $V$\\
\begin{eqnarray}
G_{c(s)}(i)=\left. \frac{dI_{c(s)}(i)}{dV} \right|_{V=0}\hspace{1cm} i=3,4 \label{Gcs}
\end{eqnarray}
In the following we shall describe the behavior of $G_c$ and $G_s$ in the two configurations~(C) and (S).
%%%%%%%%%%%%%%%%%%%%%%%%%%%%%%%%%%%%%%%%%%%%%%%%%%%%%%%%%%%%%%
%%%%%%%%%%%%%%%%%%%%%%%%%%%%%%%%%%%%%%%%%%%%%%%%%%%%%%%%%%%%%%
%%%%%%%%%%%%%%%%%%%%%%%%%%%%%%%%%%%%%%%%%%%%%%%%%%%%%%%%%%%%%%
%%%%%%%%%%%%%%%%%%%%%%%%%%%%%%%%%%%%%%%%%%%%%%%%%%%%%%%%%%%%%%
%%%%%%%%%%%%%%%%%%%%%%%%%%%%%%%%%%%%%%%%%%%%%%%%%%%%%%%%%%%%%%
\subsection{Charge bias configuration (C)}
\label{Sec4-c}
\begin{figure}[h]
\centering
%{Gc-phiFP_gamVAR_tau01}  
\includegraphics[width=\linewidth]{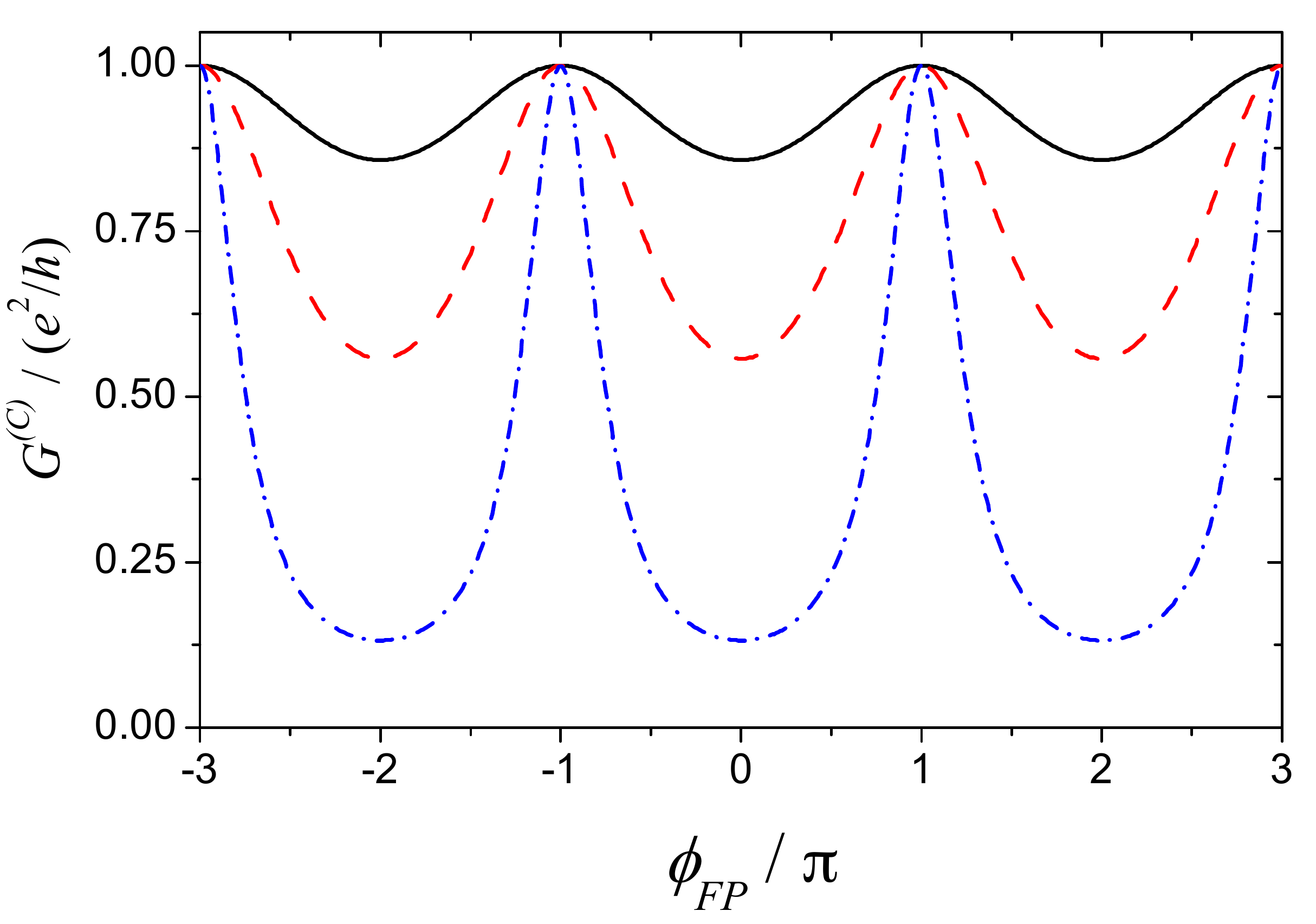} 
\caption{\label{Gc-Fig-phiFP} (Color on-line) Charge bias configuration~(C) [see Eq.(\ref{conf-c})]. The conductance (\ref{G-C}) is plotted as a function of the Fabry-P\'erot phase $\phi_{FP}= e(V_{g,T}+V_{g,B})L/\hbar v_F$,  for different values of single QPC `horizontal' transmission, namely $T^{a}_{12}=0.04$  (solid curve), $0.15$ (dashed curve) and $0.48$ (dashed-dotted curve), corresponding to spin-preserving tunneling amplitudes $\gamma_p=0.1, 0.2, 0.4$, respectively, and to spin-flipping amplitude $\gamma_f=0.1$. The Fabry-P\'erot interference pattern allows to   switch from the  `off' state (minima) to the `on' state (maxima) by operating with the gate voltages. In configuration~(C) the conductance is independent of the Aharonov-Bohm phase $\phi_{FP}= e(V_{g,T}-V_{g,B})L/\hbar v_F$.}
\end{figure}

Let us start by describing the charge bias configuration~(C)   Eq.(\ref{conf-c}).  In this case, using the expressions (\ref{Ic-gen})-(\ref{Is-gen}) for the currents, the definition~(\ref{Gcs}), and the properties~(\ref{S-prop}) of the Scattering Matrix, one obtains
\begin{equation}
\left. \begin{array}{r}
G_c(3)  \\ \\
G_s(3)  \\ \\
G_c(4)  \\\\
-G_s(4)  
\end{array}
\right\} = G^{(C)} \doteq \frac{e^2}{h} (T_{32} +T_{31} ) \label{G-C}
\end{equation}
where $T_{ij}$ denote the transmission coefficients of the setup. Using Eqs.(\ref{Tij-def}) and (\ref{Smatr-entries}), one obtains  
\begin{eqnarray}
G^{(C)} = \frac{e^2}{h}  \frac{(1-T^{a}_{12})^2}{1+T^{a}_{12} +2 T^{a}_{12}\cos{\phi_{FP}}} \label{expr-a}
\hspace{0.4cm}
\end{eqnarray}
where $T^{a}_{12}$ is  the transmission coefficient  of each single QPC defined in Eqs.(\ref{Tij1-def}), and $\phi_{FP}$ is the Fabry-P\'erot phase. Here we have assumed QPC with equal parameters, i.e. $T^{a}_{12} =T^{b}_{12}$. The oscillatory behavior of $G^{(C)}$ is shown in Fig.\ref{Gc-Fig-phiFP} for different values of $T^{a}_{12}$. Recalling the expression  (\ref{FP-phase}) for~$\phi_{FP}$, one can see that 
the oscillations have a period $\Delta (eV)= h v_F /L$ in the sum $V_{g,B} +V_{g,T}$. The maxima occur  at $\phi_{FP}=(2m+1) \pi$ ($m$ any integer), i.e. when the two paths of the FP loop [Fig.\ref{interf}(a)] interfere destructively, favouring direct transmission from terminal 2 to 3. These maxima are indeed resonances, for the tunneling amplitudes of the two QPC are equal; a slight  difference in the parameters would make resonances become  maxima very close to 1. This type of behavior of $G^{(C)}$ for the charge bias  configuration~(C) is quite similar to the gate-induced Fabry-P\'erot oscillations in the linear conductance of a   single channel quantum wire\cite{pugnetti_2009} or a  carbon nanotube transistor.\cite{liang,recher,balents}  
At the level of interference processes, the relations between such systems and the electron loops originating from spin-preserving tunneling processes have been pointed out in Section~\ref{Sec-FP}. Here we find that, in   configuration~(C), the analogy extends to the   conductance behavior as well. Indeed  in configuration~(C) the  electrodes 1 and 2 are characterized by the same Fermi distribution, and so are terminal 3 and 4. This configuration is thus topologically  very similar to a  two-terminal setup realized with carbon nanotubes   or   semiconductor quantum wires, and the dependence of $G^{(C)}$ on the Fabry-P\'erot phase $\phi_{FP}$ can thus be straightforwardly understood in terms of  such  analogy. 

On the other hand, Eq.(\ref{expr-a}) has been obtained also taking into account spin-flipping tunneling terms. In view of the discussion in Section~\ref{Sec3B}, one would thus expect a dependence of $G^{(C)}$ on the Aharonov-Bohm phase $\phi_{AB}=e(V_{g,T}-V_{g,B})L/\hbar v_F$ as well. Indeed such dependence does occur in each of the transmission coefficients $T_{31}$ and $T_{32}$ appearing in  the expression~(\ref{G-C}) for $G^{(C)}$ [see Eqs.(\ref{Tij-def}) and (\ref{Smatr-entries})].  However, the result Eq.(\ref{expr-a}) shows that this is not the case  for the sum $T_{32}+T_{31}$ and for  the  conductance $G^{(C)}$, which are independent of  $\phi_{AB}$. To understand this effect, one can consider Fig.\ref{interf} and   realize that any AB loop process  increasing the transmission from terminal  1 to terminal  3 has a partner AB loop process decreasing the transmission from  2 to 3, characterized by the opposite AB phase dependence, and leading to a perfect cancellation of the $\phi_{AB}$ dependence in the sum $T_{31}+T_{32}$.  
Notice that, nevertheless, spin-flipping tunneling amplitude $\gamma_f$ do enter Eq.(\ref{expr-a}) through the coefficients $T^{a}_{12}$ of each single QPC. One can thus conclude that, for the charge bias configuration~(C), spin-flipping processes affect  quantitatively the transmission of each single QPC, but do not lead to any qualitatively  visible interference effect between the two QPCs in the   conductance $G^{(C)}$.  We emphasize that this lack of dependence of $\phi_{AB}$ only arises because in configuration~(C)  terminals  (1,2) are biased with the equal voltages and terminals   (3,4) are grounded. It is thus not an intrinsic property of the Scattering region, but a specific feature of the charge bias configuration~(C). As we shall see, the behavior is different for the spin bias configuration~(S). \\

%%%%%%%%%%%%%%%%%%%%%
%%%%%%%%%%%%%%%%%%%%%

%%%%%%%%%%%%%%%%%%%%%%%%%%%%%%%%%%%%%%%%%%%%%%%%%%%%%%%%%%%%%%
%%%%%%%%%%%%%%%%%%%%%%%%%%%%%%%%%%%%%%%%%%%%%%%%%%%%%%%%%%%%%%
%%%%%%%%%%%%%%%%%%%%%%%%%%%%%%%%%%%%%%%%%%%%%%%%%%%%%%%%%%%%%%
%%%%%%%%%%%%%%%%%%%%%%%%%%%%%%%%%%%%%%%%%%%%%%%%%%%%%%%%%%%%%%
%%%%%%%%%%%%%%%%%%%%%%%%%%%%%%%%%%%%%%%%%%%%%%%%%%%%%%%%%%%%%%
\subsection{Spin bias configuration (S)}
\label{Sec4-s}
Let us now consider the spin bias configuration~(S)  Eq.(\ref{conf-s}).  Making again use of Eqs.(\ref{Ic-gen}), (\ref{Is-gen}), (\ref{Gcs}), and~(\ref{S-prop}), one obtains in this case
\begin{equation}
\left. \begin{array}{r}
G_c(3)  \\ \\
G_s(3)  \\ \\
-G_c(4)  \\\\
G_s(4)  
\end{array}
\right\} = G^{(S)} \doteq \frac{e^2}{h} (T_{32} -T_{31} ) \label{G-S}
\end{equation}
where $T_{ij}$  are the setup transmission coefficients defined in (\ref{Tij-def}). 
Utilizing the parametrization in terms of the single QPC transmission~(\ref{Tij1-def}), one finds 
\begin{eqnarray}
G^{(S)}=\frac{e^2}{h}     \frac{(T^{a}_{13}-T^{a}_{14})^2-4 T^{a}_{13} T^{a}_{14}  \cos{\phi_{AB}}}{1+T^{a}_{12} +2 T^{a}_{12}\cos{\phi_{FP}}}  \quad.
\label{Gstot-ii}
\end{eqnarray} 
%%%%%%%%%%%%%%%%%%%%%
\begin{figure}[h]
\centering
%{Gs-phiFP-phiAB_gam04_tau02}  
\includegraphics[width=1.0\linewidth]{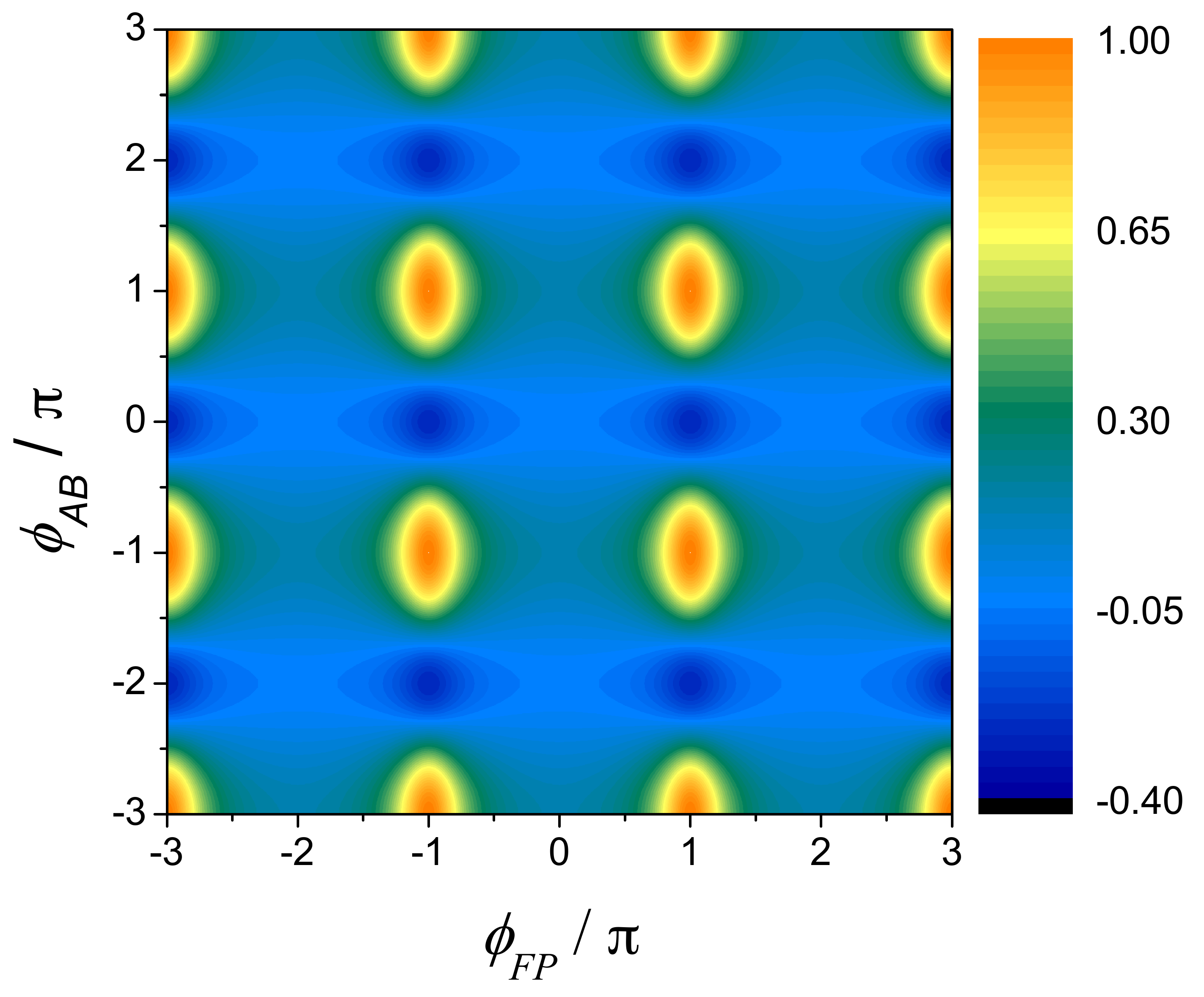}
\caption{\label{Gs-Fig-3D} (Color online) Spin bias configuration~(S) [see Eq.(\ref{conf-s})]. The conductance $G^{(S)}$ (in units of $e^2/h$) is plotted as a function of the Fabry-P\'erot phase $\phi_{FP}= e(V_{g,T}+V_{g,B})L/\hbar v_F$  and   the Aharonov-Bohm phase $\phi_{AB}=e(V_{g,T}-V_{g,B}) L/\hbar v_F$. The tunneling amplitudes for spin preserving and spin-flipping processes are $\gamma_p=0.4$ and $\gamma_f=0.2$. By tuning the gate voltages one can control both magnitude and sign of the conductance. Negative conductance values originate from spin-flipping inter-boundary coupling.
 }
\end{figure}
%%%%%%%%%%%%%%%%%%%%%
%%%%%%%%%%%%%%%%%%%%%
\begin{figure}[h]
\centering  
%{Gs-phiAB_gam04_tau02_phiFPVAR}
\includegraphics[width=\linewidth]{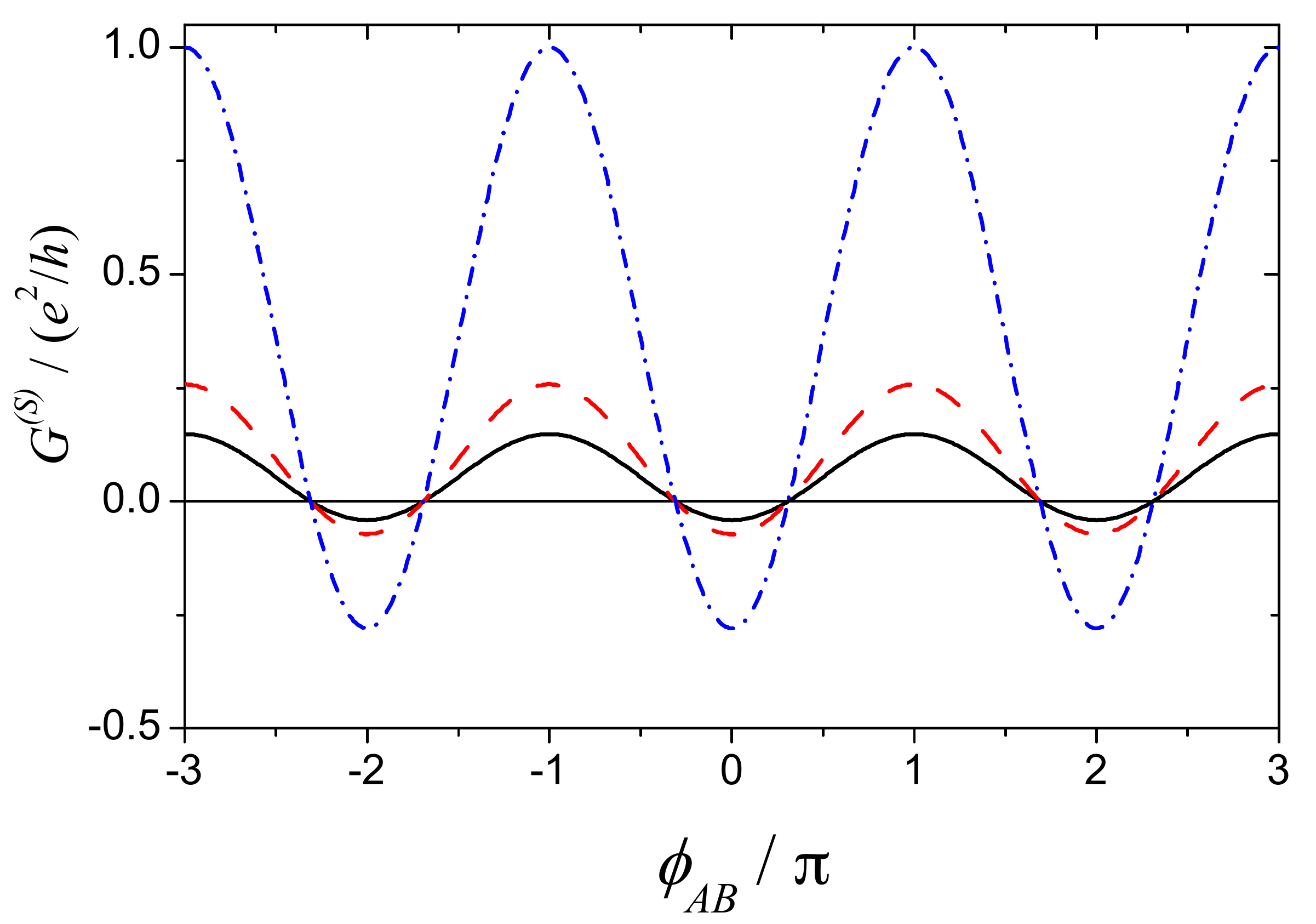}
\caption{\label{Gs-Fig-phiFP} (Color on-line) Slice plots of Fig.\ref{Gs-Fig-3D}.  The conductance  (\ref{G-S}) in the spin bias configuration~(S) is plotted as a function of the Aharonov-Bohm phase $\phi_{AB}$,  for different values of the Fabry-P\'erot phase  $\phi_{FP}$, namely $\phi_{FP}=0$ (solid curve), $\pi/2$ (dashed curve) and $\pi$ (dash-dotted curve). The dependence of $G^{(S)}$ on $\phi_{AB}$ results in an Aharonov-Bohm  modulation of the $\phi_{FP}$-dependent Fabry-P\'erot pattern, and regions of negative conductance arise.}
\end{figure}
%%%%%%%%%%%%%%%%%%%%%

\noindent Remarkably, a comparison between Eq.(\ref{G-S}) and Eq.(\ref{expr-a}) allows to realize that, while the  conductance $G^{(C)}$  obtained for the charge bias configuration~(C) only depends on the Fabry-P\'erot phase  $\phi_{FP}=e(V_{g,T}+V_{g,B}) L/\hbar v_F$,  the   conductance $G^{(S)}$ in the spin bias configuration~(S) depends on both $\phi_{FP}$ and the Aharonov-Bohm $\phi_{AB}=e(V_{g,T}-V_{g,B}) L/\hbar v_F $. The latter dependence leads to novel features in the TI setup Fig.\ref{setup-Fig} with respect to carbon nanotube based  interferometers. The behavior of $G^{(S)}$ is shown in Fig.~\ref{Gs-Fig-3D} as a function of $\phi_{FP}$ and  $\phi_{AB}$. By varying~$\phi_{FP}$, the  conductance exhibits a Fabry-P\'erot pattern, qualitatively similar to  the case of the charge bias configuration~(C)  shown in Fig.~\ref{Gc-Fig-phiFP}. However,  by varying the Aharonov-Bohm phase $\phi_{AB}$ related to gate voltage difference, one obtains a
modulation  the Fabry-P\'erot peaks, in amplitude and {\it sign}. One can thus even realize a   sign  reversal of the conductance $G^{(S)}$. This effect, absent in the conductance $G^{(C)}$, is highlighted in  Fig.\ref{Gs-Fig-phiFP}, where slices of Fig.\ref{Gs-Fig-3D} at fixed $\phi_{FP}$ values are shown, and regions of positive and  negative  conductance are clearly visible.  \\

In order to understand the physical meaning of the sign reversal and negative conductance, we first analyze the problem on the point of view of charge currents. By setting the voltage biases in configuration~(S) [Eq.(\ref{conf-s})] two charge currents are injected, one rightwards along the Top boundary and one leftwards along the Bottom boundary. 
The presence of QPCs induces inter-boundary coupling and    modifies these currents.  In particular, inter-boundary {\it forward} tunneling from one boundary opposes to the charge current counter-flowing in the other boundary, and may cause the complete blocking of the current (i.e. $G^{(S)}=0$) or even its reversal with respect to the applied voltage bias, giving rise to negative conductance values. 
This is illustrated by Eq.(\ref{G-S}), where the conductance $G^{(S)}$ is expressed as a  {\it difference}  between $T_{32}$ and $T_{31}$. Negative  conductance $G^{(S)}$  occurs whenever the `crossing' transmission $T_{31}$ (describing inter-boundary   forward  tunneling) overcomes  the `horizontal' transmission $T_{32}$ in magnitude\cite{nota-rev}. This  effect is in principle present also in a system of two usual quantum wires in presence of inter-wire tunneling.
In TI edge states, however,  helicity implies that forward tunneling processes necessarily involve a spin-flip [see Eq.(\ref{Htun-sf})], thus transferring the effect onto the spin currents as well. Notice that in the spin bias configuration~(S)  the counter-flowing charge currents correspond  to {\it parallel}   spin currents, since the charge current injected along the Top boundary   is carried by an excess of spin-$\uparrow$ right-movers, and the charge current injected along the Bottom boundary by a depletion of  spin-$\downarrow$ right-movers [see Eq.(\ref{Is-def})]. Thus, in terms of spin currents,    the tunneling term (\ref{Htun-sf}) effectively  tends to anti-align their directions. This pictorial way also explains why negative conductance are not observed in configuration~(C), where the injected spin currents are already counter-flowing. In fact, the two AB loop processes that mutually cancel in configuration~(C) (see Section \ref{Sec4-c}),  contribute with the same sign in configuration~(S).\\

 To conclude this section, we wish to emphasize the difference between our   result of negative conductance and the one obtained in Ref.[\onlinecite{chamon_2009}] for a corner junction in a 	quantum spin Hall system. In that case the possibility that a current can flow out from the lead with the lowest applied voltage   is an effect due  to   the strong electronic interaction.   In contrast, our result shows that negative conductance values can  be obtained also in the absence of electron-electron interaction, due to spin-flipping tunneling terms.   
Furthermore, we observe that  in principle  negative conductance  may   occur also in  setups with one single QPC, like the ones considered in  Refs. [\onlinecite{teo}] and [\onlinecite{trauz-recher}]. However, 
with one single QPC  the value of $T_{31}$ may not be easily  tuned: due to the linear spectrum of TI edge states, application of a gate does not lead to electron confinement,  quite similarly to what occurs in graphene. The QPC parameters are thus essentially determined at fabrication level. In contrast,  the presence of {\it two} QPCs gives rise to loop processes, opening the possibility to modulate $T_{31}$ by interferometry instead of confinement. Indeed the total `crossing' transmission $T_{31}$ depends on both the single QPC `crossing' transmissions $T^{a}_{31},T^{b}_{31}$ and on the AB phase $\phi_{AB}$, allowing  to control the sign reversal. 
Remarkably, sign reversal  only occurs as a function of~$\phi_{AB}$ and not when varying~$\phi_{FP}$. An intuitive argument to explain this property is based on the loop trajectories depicted in Fig.\ref{interf}:  as observed in Section \ref{Sec3C}, to lowest order in the tunneling amplitudes, $T_{31}$ does not depend on Fabry-P\'erot processes. This property holds to all orders, though.  
 
%%%%%%%%%%%%%%%%%%%%%%%%%%%%%%%%%%%%%%%%%%%%%%%%%%%%%%%%%%%%%%%%%%%%%%%%%%%%
%%%%%%%%%%%%%%%%%%%%%%%%%%%%%%%%%%%%%%%%%%%%%%%%%%%%%%%%%%%%%%%%%%%%%%%%%%%%
%%%%%%%%%%%%%%%%%%%%%%%%%%%%%%%%%%%%%%%%%%%%%%%%%%%%%%%%%%%%%%%%%%%%%%%%%%%%
%%%%%%%%%%%%%%%%%%%%%%%%%%%%%%%%%%%%%%%%%%%%%%%%%%%%%%%%%%%%%%%%%%%%%%%%%%%%
%%%%%%%%%%%%%%%%%%%%%%%%%%%%%%%%%%%%%%%%%%%%%%%%%%%%%%%%%%%%%%%%%%%%%%%%%%%%
%%%%%%%%%%%%%%%%%%%%%%%%%%%%%%%%%%%%%%%%%%%%%%%%%%%%%%%%%%%%%%%%%%%%%%%%%%%%
\section{Discussion}
\label{Sec5}
The results presented in the previous section show that in both the `Charge'(C) and `Spin'(S) bias configurations   a tuning of the   conductance   is possible   operating with the gate voltages $V_{g,T}$ and $V_{g,B}$. We recall that $G^{(C)}$ and $G^{(S)}$ represent (up to a sign) the linear response of charge and spin currents in terminals 3 and~4 [see Eqs.(\ref{G-C}) and (\ref{G-S})]. In particular, this implies that the interferometer Fig.\ref{setup-Fig} allows for an all-electric tuning of the spin current, without invoking magnetic fields or polarized ferromagnets. As observed in the introduction, this feature represents an advantage in view of   miniaturization with respect to spin-based devices involving magnetic materials, due to the difficulty in realizing  magnetic fields at nanoscale resolution.  Furthermore, due to intrinsic absence of backscattering in TI edge states,   the  interferometer considered here exhibits  typical values of the conductance maxima   of the order of the conductance quantum, a relevant aspect for applications as well. In spintronics devices based on ferromagnet-normal metal hybrid  junctions, for instance, the high tunnel energy barriers severely limit transistor conductance in the `on' state,   reducing the  current delivery capability per channel.\\
We now wish to  comment the specific results obtained for each configuration. \\

In charge bias configuration~(C)   Eq.(\ref{conf-c}) the  conductance~$G^{(C)}$ exhibits Fabry-P\'erot oscillations as a function of the FP phase $\phi_{FP}= e(V_{g,T}+V_{g,B})L/\hbar v_F$ similar to the case of nanotubes and quantum wires. Notably, in this configuration charge and spin conductances are independent of the Aharonov-Bohm phase $\phi_{AB}$, so that the additional degree of freedom  $V_{g,T}-V_{g,B}$ provided by the edge channel space separation with respect to the case of carbon nanotubes is actually ineffective. In this respect configuration~(C) is topologically very similar to the case of other two-terminal charge-based electron interferometers. 
Indeed  Eqs.(\ref{G-C}) imply  that, if one merges the currents flowing into terminals 3 and 4,  a {\it pure  charge} signal is obtained, i.e.
\begin{eqnarray}
G^{tot}_{c} \doteq G_{c}(3)+ G_{c}(4)&=& 2 \, G^{(C)}  \label{Gtotc-i} \\
G^{tot}_{s} \doteq G_{s}(3)+ G_{s}(4) &=& 0 \quad .
\end{eqnarray}
in accordance with the two-terminal setup analogy.
Nevertheless,  because of the properties of TI edge states, the four terminal setup  offers  the advantage of performing  spin-resolved measurements, since currents flowing in terminals 3 and 4 are carried by a   specific majority spin component.  Indeed configuration (C) can for instance be exploited to investigate charge and spin tunneling currents through one QPC, as shown in Ref.[\onlinecite{strom_2009}], where relations similar to (\ref{G-C}) have been found between $G_{c}$ and $G_{s}$ even in the presence of interaction.\\

A richer scenario is obtained for the spin bias configuration~(S) [see Eq.(\ref{conf-s})], intrinsically different from a two-terminal setup. In this case the Fabry-P\'erot oscillations of $G^{(S)}$ driven by~$\phi_{FP}$ are modulated by the Aharonov-Bohm oscillations driven by $\phi_{AB}= e(V_{g,T}-V_{g,B})L/\hbar v_F$.
In particular a sign reversal of the conductance is possible by varying the gate voltage difference.  
As observed in Section \ref{Sec4-s}, the sign reversal of the conductance is caused by the spin-flipping terms (\ref{Htun-sf}) that effectively behave as an  anti-aligning coupling for the injected spin currents.
Furthermore,  we point out that in the bias configuration~(S) the charge conductances in terminals 3 and 4 have equal magnitude but opposite signs, whereas the spin conductances have equal magnitudes and signs. By merging the currents flowing in terminals 3 and 4
\begin{eqnarray}
G^{tot}_{c} \doteq G_{c}(3)+ G_{c}(4)&=& 0 \label{Gtotc-ii} \\
G^{tot}_{s} \doteq G_{s}(3)+ G_{s}(4) &=& 2 \, G^{(S)}  \label{Gtots-ii} 
\end{eqnarray} 
one obtains  a {\it pure spin}   current,  a flow of electronic angular momentum that is not accompanied by any net charge current.\cite{awschalom-science-04} 
The realization of pure spin currents represents an extremely attractive problem nowadays, for it allows to avoid charge-related spurious decoherence effects, and possibly opens new perspectives for coherent transport and quantum computing. Indeed a number of schemes  have been proposed to generate pure spin currents.\cite{pure-spin-gen}   
On the other hand, the absence of any net charge current also makes the detection of pure spin currents a difficult task, which requires ad hoc methods exploiting e.g. optical techniques\cite{long,Zhao} or spin-valve effect.\cite{spin-valve} The setup described here, based on TI edge states, is relatively versatile in this respect. Exploiting TI edge state properties, it allows to generate spin currents by setting the voltage biases in configuration~(S), to measure them with ordinary methods (amperometers) in separate terminals 3 and 4, and to adjust the conductances to the desired values by operating with the gate voltages. Then,   by merging the signals of terminals 3 and 4, a pure spin current is obtained. The ballistic behavior yields values of conductances comparable with $e^2/h$.  These aspects are expected to contribute to make the use of spin currents a realistic perspective within a short time.\\

We now  provide some typical values of the relevant quantities involved in the setup, which   can be realized in  HgTe quantum wells\cite{molenkamp-science}, where multi-terminal transport measurements have recently been performed\cite{molenkamp-science-2}. The period characterizing the oscillatory behavior of the conductances   as a function of $\phi_{FP}$ and $\phi_{AB}$, when expressed in terms of gate voltages, equals  $e\Delta V_{g,T/B}=h/v_{F} L$. Assuming a distance between the two QPCs $L \simeq 1 \,{\rm \mu m}$, and a Fermi velocity\cite{trauz-recher} $v_F \simeq 0.5 \cdot 10^6 {\rm m/s}$, one obtains a value $e\Delta V_{g,T/B} \simeq 1 \, {\rm meV}$. 
The possibility for conductance sign reversal is related to the order of magnitude of $T_{31}$, the `crossing' transmissions of the setup. The important parameter that ultimately determines
$T_{31}$ is the value of the single QPC transmission $T^{a}_{31}$ and $T^{b}_{31}$. We emphasize that relatively weak spin-flipping probabilities are sufficient to obtain sign reversal. Indeed, the values $\gamma_p=0.4$ and $\gamma_f=0.2$ used for Fig.\ref{Gs-Fig-3D} correspond to single QPC `crossing' transmissions  $T^{a}_{31}=T^{b}_{31}=10 \%$ and $T^{a}_{32}=T^{a}_{41}=T^{b}_{32}=T^{b}_{41}=45 \%$.
Assuming two equal QPCs,  one can see from Eq.(\ref{Gstot-ii}) that the condition to observe a sign reversal can be expressed as  the inequality
\begin{equation}
(T^{a}_{31}-T^{a}_{14})^2-4 T^{a}_{31}  T^{a}_{14}  < 0
\end{equation} 
in terms of the single QPC transmission probabilities.
The values of $T^{a}_{31}$ and $T^{a}_{14}$ can be operatively determined by biasing only terminal 1 ($V_1=V$ and $V_2=V_3=V_4=0$) and measuring the linear response of the currents in terminals 3 and 4, respectively. \\

Before concluding, a comment about the effect of electron-electron interaction is in order. Similarly to other one-dimensional systems like carbon nanotubes or semiconductor quantum wires, in TI edge states  the Coulomb interaction, which is screened at short and long distances, leads to a Luttinger liquid behavior\cite{zhang-PRL,chamon_2009,trauz-recher,strom_2009,moore-PRB-2006,ojanen_2010}. For edge states in HgTe quantum wells the Luttinger parameter $g$, describing the strength of the screened Coulomb interaction, has been estimated to range from the weakly interacting limit $g \simeq  1$~\cite{maciejko_2009}  to the moderate interaction regime $g \sim 0.8$~\cite{teo,strom_2009}, down to the strongly interacting case $g \sim 0.5$~\cite{chamon_2009}, depending  on both geometrical parameters  such as the quantum well thickness  and the conditions of material growth. The analysis carried out here   has focussed on the regime of non-interacting edge states. This has enabled us to derive  a complete solution of the edge state dynamics for arbitrary values of tunneling parameters and gate voltages, which we expect to qualitatively hold for weak interaction too. In contrast, in the regime of strong interaction  the Scattering Matrix approach is not applicable and an exact solution is not available. The use of other methods (such as Renormalization Group Analysis or perturbative treatments) is   mandatory\cite{zhang-PRL,moore-PRB-2006,strom_2009}, and determining the properties of the setup in the strongly interacting case represents a demanding task in general. For a single QPC or a corner junction, for instance, it has been shown that Luttinger liquid signatures emerge as a power-law behavior in transport properties~\cite{chamon_2009,strom_2009,trauz-recher,ojanen_2010}. For the two QPC interferometric setup in Fig.\ref{setup-Fig}, one can speculate about interaction effects exploiting some analogies with recently studied Fabry-P\'erot interferometers in the strongly correlated regime\cite{pugnetti_2009}. In particular, a modification of the oscillation periods in~$\phi_{FP}$ and $\phi_{AB}$ may be expected. Also, in the low bias regime the non-interacting limit result may qualitatively be preserved by the finite distance between the electrodes, while Luttinger liquid signatures should emerge at non-linear transport. However, we emphasize that   the helical nature of edge states in Topological Insulators makes them intrinsically different from other Luttinger liquid realizations, possibly causing the emergence of new features. Furthermore, strong correlation  makes edge states more sensitive to local defects such as fluctuations of ion concentration in the doping layers\cite{sherman,gui} or random bonds at the quantum well interfaces\cite{golub} that may even localize the edge states\cite{strom_2010}. For these reasons the determination of transport properties of setup Fig.\ref{setup-Fig} in the strongly interacting regime is an extremely rich and interesting problem, which deserves a separate analysis beyond the purpose of the present paper. 

\section{Conclusions}
\label{Sec6}
In conclusion, we have shown that interferometry of TI edge states allows to realize a full electrically controllable charge and spin transistor, without the use of magnetic fields. The charge and spin conductances of the proposed setup in Fig.\ref{setup-Fig} can be tuned by operating with electric gate voltages applied to the Top and Bottom regions between the two QPCs, which modify the phase of the two interference phenomena characterizing the system:   Fabry-P\'erot-like loops generated by spin preserving tunneling processes and  Aharonov-Bohm-like loops generated by spin flipping tunneling processes. 
Two voltage bias configurations have been particularly addressed [Eqs.(\ref{conf-c})-(\ref{conf-s})]. The charge bias configuration~(C) leads to a conductance behavior that is qualitatively similar to the Fabry-P\'erot oscillations observed in carbon nanotubes electron interferometers, with the additional advantage of allowing for a measure of spin-resolved conductances as well. In the spin bias configuration~(S), where counter-flowing injected charge currents correspond to parallel  spin currents, the  Fabry-P\'erot conductance pattern is modulated by Aharonov-Bohm oscillations. This modulation originates from spin-flipping tunneling processes, tending to anti-aling the spin currents and leading to a current blocking or reversal with respect to the applied bias, yielding negative conductance values. Furthermore, this configuration also enables to obtain a pure spin current.
These peculiar features suggest that the proposed TI edge state setup may have a relevant interest for spintronics applications.  
%%%%%%%%%%%%%%%%%%%%%%%%%%%%%%%%%%%%%%%%%%%%%%%%%%%%%%%%%%%%%%%%%%%%%%%%%%%%%%%%%%%%
%%%%%%%%%%%%%%%%%%%%%%%%%%%%%%%%%%%%%%%%%%%%%%%%%%%%%%%%%%%%%%%%%%%%%%%%%%%%%%%%%%%%
%%%%%%%%%%%%%%%%%%%%%%%%%%%%%%%%%%%%%%%%%%%%%%%%%%%%%%%%%%%%%%%%%%%%%%%%%%%%%%%%%%%%
%%%%%%%%%%%%%%%%%%%%%%%%%%%%%%%%%%%%%%%%%%%%%%%%%%%%%%%%%%%%%%%%%%%%%%%%%%%%%%%%%%%%
%%%%%%%%%%%%%%%%%%%%%%%%%%%%%%%%%%%%%%%%%%%%%%%%%%%%%%%%%%%%%%%%%%%%%%%%%%%%%%%%%%%%
%%%%%%%%%%%%%%%%%%%%%%%%%%%%%%%%%%%%%%%%%%%%%%%%%%%%%%%%%%%%%%%%%%%%%%%%%%%%%%%%%%%%
%%%%%%%%%%%%%%%%%%%%%%%%%%%%%%%%%%%%%%%%%%%%%%%%%%%%%%%%%%%%%%%%%%%%%%%%%%%%%%%%%%%%
%%%%%%%%%%%%%%%%%%%%%%%%%%%%%%%%%%%%%%%%%%%%%%%%%%%%%%%%%%%%%%%%%%%%%%%%%%%%%%%%%%%%
%%%%%%%%%%%%%%%%%%%%%%%%%%%%%%%%%%%%%%%%%%%%%%%%%%%%%%%%%%%%%%%%%%%%%%%%%%%%%%%%%%%%
%%%%%%%%%%%%%%%%%%%%%%%%%%%%%%%%%%%%%%%%%%%%%%%%%%%%%%%%%%%%%%%%%%%%%%%%%%%%%%%%%%%%
%%%%%%%%%%%%%%%%%%%%%%%%%%%%%%%%%%%%%%%%%%%%%%%%%%%%%%%%%%%%%%%%%%%%%%%%%%%%%%%%%%%%
%%%%%%%%%%%%%%%%%%%%%%%%%%%%%%%%%%%%%%%%%%%%%%%%%%%%%%%%%%%%%%%%%%%%%%%%%%%%%%%%%%%%
\acknowledgments
The author greatly acknowledges B. Trauzettel, C.-X. Liu, P.~Recher, F.~Taddei and D.~Bercioux for fruitful discussions and for their careful reading of the paper.


\begin{thebibliography}{99}
\bibitem{nota-1} We use for  spin current  the same dimensions as the charge current, instead of the electron angular momentum current $(\hbar/2) v_F$. This seems to be more appropriate in TI edge states, where charge and spin current are closely related, see also Ref.[\onlinecite{trauz-recher}].
\end{thebibliography}
\end{document}